\documentclass[journal,draftclsnofoot,onecolumn,12pt]{IEEEtran}
\usepackage{amsthm,amssymb,graphicx,multirow,amsmath,color,amsfonts}
\usepackage[update,prepend]{epstopdf}
\usepackage[noadjust]{cite}
\usepackage[latin1]{inputenc}
\usepackage{tikz}
\usetikzlibrary{angles, quotes}
\usepackage{bbm} 
\usepackage{pdfpages}
\usepackage{tabulary}
\usepackage{multirow}
\usepackage{comment}
\usetikzlibrary{calc}
\usepackage{ amssymb }
\usepackage{subfig}
\usepackage{hhline}

\usepackage{bm}

\def\nb0{{\mathbf{0}}}
\def\nb1{{\mathbf{1}}}

\allowdisplaybreaks 
\usepackage{setspace}    

\usepackage{gensymb}

\begin{document}
\graphicspath{{./Figures/}}
\title{Exploiting Wind Turbine-Mounted Base Stations 
to Enhance Rural Connectivity}
\author{
Maurilio Matracia, \em Student Member, IEEE, \normalfont Mustafa A. Kishk, \em Member, IEEE, \\ \normalfont and Mohamed-Slim Alouini, \em Fellow, IEEE 
\thanks{The authors are with Computer, Electrical and Mathematical Science and Engineering (CEMSE) Division, King Abdullah University of Science and Technology (KAUST), Thuwal 23955-6900, Kingdom of Saudi Arabia (KSA) (email: $\{$maurilio.matracia; mustafa.kishk; slim.alouini$\}$@kaust.edu.sa). 
}
\vspace{-2mm}}

\maketitle

\begin{abstract}
Although global connectivity is one of the main requirements for future generations of wireless networks driven by the United Nation's Sustainable Development Goals (SDGs), telecommunication (telecom) providers are economically discouraged from investing in sparsely populated areas, such as rural and remote ones. 
Novel affordable and sustainable paradigms are thus indispensable to enhance the cellular infrastructure in such areas and bridge the digital divide when compared with urban ones. 
We investigate the use of wind turbine-mounted base stations (WTBSs) as a cost-effective solution for regions with high wind energy potential, since it could replace or even outperform current solutions requiring additional cell towers (CTs), satellites, or aerial base stations (ABSs).
Indeed, conveniently installing base station (BS) equipment on wind generators would allow the transceivers to reach sufficient altitudes and easily establish line-of-sight (LoS) channels within large areas. 
We also propose insightful simulation results for realistic case studies based on data sets for wind speed and population densities as well as wind turbines' (WTs') and CTs' locations within specific French, Argentine, and Ethiopian exurban regions.
By doing this, we hope to prove the feasibility and the effectiveness of this solution and stimulate its implementation.
\end{abstract}

\begin{IEEEkeywords}
Digital divide, wind turbines, cell towers, sustainability.
\end{IEEEkeywords}

\section{Introduction} \label{sec:Intro}
Vast under-connected exurban regions lacking a reliable Internet and communication technology (ICT) infrastructure are still disseminated worldwide \cite{Yaacoub19,zhang21telecommunication}. 
Therefore, a primary humanitarian goal for our society is to reduce the digital gap between these regions and the more advanced ones, which are mostly urban.
Achieving this goal will in turn enable many important business opportunities and technologies for rural users, but the conventional solution of deploying CTs inherently raises at least three important issues: \\
(i) The typical height of these structures is limited to a few tens of meters, resulting in relatively small coverage radius (a few kilometers) since the probability of establishing LoS links at larger distances becomes insufficient.
This eventually leads to an excessive cost, since a very capillary infrastructure is needed to cover the area of interest;\\
(ii) Generally speaking, CTs' visual impact strongly disturbs rural inhabitants (note that we exclude the option of concealed CTs due to their excessive cost for a typical rural economy).
Due to this, indeed, many countries worldwide have established strict regulations on their deployment;\\
(iii) Since the majority of rural areas lack ubiquitous and reliable power grids, it is often required to install diesel generators to serve either as a main or a back-up energy source for the telecom infrastructure.
However, diesel generators are not sustainable nor autonomous since they produce large amounts of pollutants and require continuous fuel refilling, respectively.\par
Based on the aforementioned considerations, an effective alternative could be to mount BS equipment on WT towers to take advantage of their tall and strong structure, which is usually already connected to the power grid and even the telecom network, for transmitting data (e.g., temperature, wind speed, humidity) to its diagnostic/control center, which in turn monitors and controls the WT itself. \par
The idea of providing communication functionality to WT towers was first conceived by Thomas Michael Sievert in \cite{Patent}, but to our best knowledge it has not been implemented yet for cellular coverage enhancement.
The dual solution is to mount a small WT on a CT (horizontal axis WTs have been mounted on top or aside CTs, whereas vertical axis ones can be hosted even inside lattice structures), possibly introducing a secondary source that ensures continuous power supply.
In such case, however, the height of the transceiver is still very limited and the problem of transmitting in LoS condition persists.  \par
The rest of this paper is organized as follows.
While Sec. \ref{sec:Intro} concludes with a brief overview of the related works, the next section is an overview of the main aspects concerning the improvement of the quality of service (QoS) in rural areas.
Then, the proposed solution is introduced and discussed from various perspectives in Sec. \ref{sec:WTBS}.
The importance of the planar distributions of wind speed and population densities as well as the locations of the CTs and the WTBSs is discussed in Sec. \ref{sec:distr}.
Finally, Sec. \ref{sec:proof} provides realistic case studies with insightful simulation results and Sec. \ref{sec:conclusion} concludes the paper.

\vspace{-1mm} \subsection{Related Works}

\vspace{1.5mm}       \subsubsection{Coverage Enhancement in Rural Areas}
Recently, researchers have suggested several options to provide better services to rural users.
A comprehensive overview of the most important fronthaul and backhaul paradigms for rural communications has been provided in \cite{Yaacoub19}.
One of the most promising paradigms consists in supporting terrestrial base stations (TBSs) with ABSs, and its effectiveness in improving the access links has been evaluated in \cite{Matracia20rural}. 
Satellite communications systems have also gained increasing attention \cite{Talgat2020}, but are usually more appropriate for very large rural regions lacking backhaul links.  \par

\vspace{1.5mm}      \subsubsection{Wind Energy Harvesting} \label{subsubsec:WEH}
In the context of energy harvesting, WTs represent one of the main pillars.
Extensive overviews of the types of generators used with wind energy conversion systems as well as the emerging technologies have been provided in literature works such as \cite{WindReview}.
Authors in \cite{Ahmed14smartWF} have also discussed the networking of WTs in wind farms (WFs), but they did not consider the possibility of enhancing the QoS experienced by nearby users. \par 
Note also that many innovative designs have been proposed for harvesting energy in more efficient and safe ways.
Among all, the most promising designs is probably the wind oscillator developed by the Spanish startup Vortex Bladeless. 
This is a vortex induced vibration resonant wind generator that consists of a carbon/glass fiber cylinder, fixed vertically with an elastic rod, generating electricity from its own oscillations.
Many other interesting designs have been developed for wind energy harvesting during the last decade, but few of them have been realized (e.g., the Saphonian by Saphon Energy and Invelox by SheerWind) and none of them has been commercialized on a large scale yet. 

\section{Service Enhancement in Rural Areas}
Generally speaking, rural areas are strongly under-connected compared to nearby towns.
The digital divide between urban and remote areas is often extreme in the least developed countries and still present everywhere.
For example, the percentage of African urban households with Internet access is multiple times larger than its rural counterpart, whereas the difference is minor when considering any European country.

\vspace{-1mm} \subsection{Alternative Solutions for Rural Connectivity}
Several new architectures are expected to change the near future of rural connectivity.
The most impactful and cost-effective might be the following.

\vspace{1.5mm}       \subsubsection{Facebook Connectivity's SuperCell} 
This project aims to build extremely tall CTs (up to $250\,$m) that use $36$ azimuthal sectors for higher capacity, in order to reduce the total cost of the infrastructure.
The field measurements mentioned in \cite{FBsupercell}, in fact, reveal that the coverage area of a single SuperCell could be around $65$ times larger compared to the one of a conventional macro CT ($30\,$m tall).
Thus, the required density of CTs might even be reduced by twenty times.
This tall guyed mast, however, requires several bulky ropes for stability and hence occupies a very large area.

\vspace{1.5mm}       \subsubsection{Altaeros' SuperTower}
This is an autonomous, high capacity, long-endurance tethered airship designed in 2019.
Its optional mobile bases enable a fair amount of relocation flexibility that a drone would be able to outperform only at the price of smaller coverage radius, capacity, and autonomy.
The ST-300 model can lift $300\,$kg of payload up to $300\,$m in altitude, far beyond the capabilities of tethered drones.
Harsh environmental conditions still represent a key challenge for this solution as well as for any other kind of low-altitude platform (LAP).

\vspace{1.5mm}       \subsubsection{High-altitude platforms (HAPs)}
This technology makes use of balloons, airships, or gliders equipped with transceivers which are mainly powered by solar energy.
HAPs operate within $17$-$50\,$km in altitude, thus providing either high capacity per unit area or extremely large coverage, respectively.
Although weather conditions are relatively more favorable in the stratosphere, HAPs' performances can be reduced in case of underlying dust storms. 

\vspace{1.5mm}       \subsubsection{SpaceX's Starlink}
The idea behind Satellite Internet is to provide global wireless coverage by means of a myriad of LEO satellites, and Earth stations operating as relays between users and satellites.
While Starlink and several other competitive projects are progressing, there are still two important concerns associated to this technology: 
(i) satellites can disturb astronomic observations and prevent from detecting dangerous asteroids as well as discovering of new planets or black holes, and (ii) if a satellite collides with another object, it can be shattered in thousands of debris which in turn can collide with other satellites.
SpaceX partially solved the first issue by designing DarkSat (a satellite with black antireflective coatings) first and VisorSat (which is also equipped with a sunshade) then.
On the other hand, the second issue is considered more serious since the density of LEO satellites is rapidly increasing and some of them are already adrift.

\vspace{-1mm} \subsection{Sustainability and Public Opinion}
Whenever a new technology is proposed, its environmental and societal impacts as well as other contextual factors must be taken into account.
Hereby, we briefly discuss the main concerns that governments and inhabitants express regarding the actual solutions and we try to predict what should their reaction be to the novel ones. \par
Regarding the public opinion on WTs' deployment, an interesting work has been conducted in \cite{WT_publicOpinion}, explaining why rural communities' lack of trust in commercial developers is often a limiting factor, for instance. 
Deploying CTs can be even more problematic, since it is believed that they have a stronger visual impact on the landscape and population often perceives them as more dangerous than WTs.
Several television reports have also showed the negative effect of CTs on the business of a company or a household market.

\section{The Wind Turbine-Mounted Base Station} \label{sec:WTBS}
\vspace{1mm} \subsection{Proposed Designs}
Differently from many other RESs, large wind generators can operate almost full time during the whole year since the allowed wind speed approximately ranges from $2$ to $25\,$m/s. 
Since they represent a precious resource spread worldwide in remote areas, we propose to take advantage of their robust structures to host telecom transceivers and ameliorate rural connectivity. 
Whenever already interconnected to the electrical infrastructure, WTBSs might also be extremely cost-effective, since they would reduce the need of deploying additional CTs, satellites, or ABSs. 
Mounting full BS equipment on wind generators would simply allow to reach sufficient altitudes to ensure LoS communication channels within large areas, without the need of building new structures. 
To this extent, several options can be considered: \\
$\bullet$ Given that a large WT's nacelle is approximately $3\,$m in height and width, and $8\,$m in length, resulting in a volume of $72\,$m$^3$, there might often be enough space available for mounting the BS equipment inside it, so that the BS would be concealed, easily accessible, and protected from harsh weather. 
However, the BS transmit power should be slightly increased because of the relatively small attenuation of the signal due to the fiber glass walls of the nacelle itself.
Telecom companies such as the American Telephone and Telegraph Company (AT$\text\&$T) actually make use of fiber glass as a radio frequency (RF) transparent material to conceal BS equipment in churches or other buildings; \\
$\bullet$ If the nacelle is already full and the roof of the WT is easily accessible, the BS equipment can be mounted on top of it, so that the maximum height is reached and the quality of the communication channel is optimized for the farthest users; \\
$\bullet$ If the nacelle is already full and the roof of the WT is not easily accessible, the BS equipment can be mounted along the tower itself, beneath the hub. 
In this case, the telecom equipment would be completely independent from the WT components, but the coverage radius would be smaller compared to the previous options. 
Moreover, if the transceivers are installed at an altitude below the area spanned by the blades, any possible interference (due to scattering or reflection) caused by the latter would be avoided. \par
Evidently, the first case requires the energy and the telecom operators to share the same room, while for the second and especially the third case they would be more independent from each other. 
Note that analogous considerations can be done for novel designs such as the Saphonian wind generator mentioned in Sec. \ref{subsubsec:WEH}, if its size is increased to be comparable to large WTs.
However, the higher mobility of its nacelle should be taken into account.
For the case of wind oscillators such as Vortex Bladeless, instead, the most convenient option might be to install the transceivers on top of the structure, although the effect of its oscillations should be investigated.

\vspace{-1mm} \subsection{Backhaul Solutions}
Having a reliable backhaul link is vital for ensuring sufficient QoS to users.
Fortunately, wind towers are usually connected with optical fiber to their diagnostic/control center (or to the core network, in general).
This would represent the best case for backhaul connection, although the capacity of the respective link might need to be increased for larger data traffics.
Wherever optical fiber is not already available (this usually happens because it is not economically convenient to dig channels to lay several kilometers of optical fiber underground), if the WT is connected to the grid then it might be possible to wrap the optical fiber around overhead power lines by means of Facebook's connectivity robot, Bombyx, 
or alternatively rely on point to point (P2P) backhaul by means of microwave channels since the nacelles are usually made of RF-transparent materials such as glass fiber-reinforced plastic (GFRP). \par
Another interesting option could be to possibly take advantage of the deployment of LEO satellites. 
Indeed, it might be convenient to install a satellite dish on top of the nacelle of the tallest WT of a WF and use it to connect the other ones in a star-fashion by using millimeter wave (mmWave) or television white space (TVWS) unlicensed spectrum links. 
Again, nearby rural users could access the Internet by associating to the BSs attached to any of the equipped WTs.

\vspace{-1mm} \subsection{Energy-Related Considerations}
The power consumption of the BS equipment is usually negligible compared to the power produced by a large WT. 
A typical BS equipment requires about $5$-$10\,$ kW, of which around two-thirds is due to the RF equipment and the remaining part is due to the air conditioning system, the digital signal processor, and the AC/DC converter.
Assuming a grid-connected WT, the optimal choice is usually to connect the BS equipment directly to the grid, in order to minimize the risk of interrupting  telecom services.
For stand-alone WTs, instead, a back-up source connected to the DC link of the WT's inverter (e.g., a battery pack, a fuel cell system, or a diesel generator) would be needed to ensure a continuous service, but this would probably require to place it outside the tower since the available space inside is quite limited.
Another important aspect regards the possibility of consuming green energy directly from the source, since the latter can considerably reduce both the cost and the environmental impact of any energy-intensive application.
A noteworthy example is WindCORES, a collaboration between Fujitsu's Green IT initiative and WestfalenWIND IT aiming to take advantage of the space available at the base of WT towers to host sustainable data centers at competitive prices.

\vspace{-1mm} \subsection{Economic Aspects}
When talking about rural connectivity, the largest percentage of the total cost derives from building both the power and the telecom infrastructures.
While many sparsely populated areas do not have access to any of them, some others enjoy the availability of a power grid.
However, BSs expect a reliable power supply in order to function continuously over years, thus both the capital expenditures (CAPEX) and the operating expenditures (OPEX) may become excessive.\par
The proposed solution expects to drastically reduce the cost of these infrastructures by simply avoiding to build CTs and, in some cases, even to dig optical fiber cables.
In fact, if the WT is already connected to the core network then the existing link could also support additional mobile users or devices for Internet of things (IoT) applications.
By means of their financial support, telecom providers might also be a key-enabler for deploying new WTBSs and reduce the carbon footprint of the telecom industry.
This is why several companies, including AT$\text\&$T, have started relying more and more on WTs and other renewable energy sources for powering their own infrastructures, also taking advantage of considerable government subsidies.
For existing WTs, other kinds of agreements such as a simple rent for hosting the BS equipment should be considered by power and telecom providers. \par
Finally, we also highlight that the proposed solution should not require any additional permission from the public authorities and hence it could be implemented in a much shorter time compared to the conventional ones.

\section{Planar Distributions} \label{sec:distr}
When evaluating the possibility of deploying WTBSs, very accurate plannings should consider the average distance between the tower and the interface with the power grid, since the BS equipment is preferably connected to the latter.
However, the three main entities that should be preliminarily taken into account are population, existing CTs, and existing WTs.
In this section we discuss the required features for the planar density distributions of these entities over the environment considered.\par
Population maps need to be analyzed carefully in order to understand whether the site is appropriate for equipping wind towers or not.
Evidently, if there is a scarce density of users, the cost of providing cellular coverage might not be justified, unless unmanned applications such as industrial IoT and precision farming are needed.
A minimum distance of a few hundred meters (mainly depending on the level of urbanization and the size of the WT) between any WT and any inhabited center should be also considered to meet government's regulations and avoid disturbing nearby residents. \par
Regarding existing BSs, if their density is already high within the considered area then it might not be useful to increase their number by equipping any wind towers, although this case is very rare in rural areas.
In other words, mounting transceivers on WTs would be most effective when there is a low density of CTs and a large density of users. \par
Although this solution is also applicable to new wind towers, its economic effectiveness should be maximized by taking advantage of existing structures.
Trivially, the more wind towers are already available, the more choices we have about which of them should be equipped to maximize network's performances.
Note that countries such as Denmark already dispone of an average density of onshore WTs that is theoretically enough to cover all their rural communities, even without the support of existing CTs.

\section{Proof of Concept} \label{sec:proof}
This section provides interesting  simulation results showing the performance enhancements that proper deployments of WTBSs could bring to existing networks. \par
Since 5G CTs are not available yet in the considered exurban areas \cite{opencellid}, we assume WTBSs to operate as 4G BSs.
Furthermore, given that the data sets in \cite{data_Wpower,opsd} refer to the locations of the center of the WFs, we will assume to mount at most one BS equipment per WF, which will also minimize the interference due to clustered BSs as well as the costs of WTBSs' deployment.

\vspace{-1mm} \subsection{Channel Model}
Our system model conceives six possible types of communication links, depending on the structure hosting the BS (CT $T$ or WTBS $W$), the mobile technology (3G or 4G for CTs, 4G for WTBSs) and the type of transmission (LoS $L$ or NLoS $N$) with respect to the user.
Let the subscript $Q\in\{T\!L 3,T\!N 3,T\!L 4,T\!N 4,
W\!L 4,W\!N 4\}$ denote the type of BS, with the first letter referring to the structure, the second to the type of transmission, and the number referring to the generation of mobile technology. 
We assume that each wireless link between the user and the BS $Y_i$ experiences small-scale fading in the form of Nakagami-m distribution with its specific shape parameter $m_Q$, and channel fading power gain $G_{Q,Y_i}$ following a Gamma distribution.
We also introduce the mean additional losses $\eta_Q$'s and the transmit powers $p_T$ and $p_W$ and consider a standard power-law path-loss model characterized by the path-loss exponents $\alpha_Q$'s. \par
Note that NLoS and LoS transmissions occur with specific probabilities depending on the height and density of the buildings, the type of environment, and the elevation angle.
Indeed, according to \cite{al2014optimal} the LoS probability as a function that depends on two environmental constants, called s-curve parameters ($\mathcal{S}_a$ and $\mathcal{S}_b$), and increases as the elevation angle of the BS increases.
Trivially, the NLoS probability is complementary to the LoS one and each BS is assumed to be either in LoS or NLoS condition with the user, independently of the other BSs. 
Finally note that, for the sake of simplicity, we neglect antenna gains and assume all the BSs to be omnidirectional.

\vspace{-1mm} \subsection{Data rate}
We adopt the strongest average received power association policy, therefore the user connects to the BS providing the highest average received power.
However, in order to maximize the average data rate per user, we penalize the association to 3G BSs by introducing a bias factor on the power that would be received in case of association to a 4G BS.
The instantaneous signal-to-interference-plus-noise ratio (SINR) will be the ratio between such received power from the serving BS and the sum of the aggregate interference (sum of the received powers from all the other BSs) and the additive white Gaussian noise (AWGN) power $\sigma_n^2$.
The coverage probability is defined as the probability that the SINR exceeds a designated threshold $\beta$, which leads to introducing the lower bound $\bm\bar{R}$ of the average data rate as the product between the coverage probability and the data rate $R_Q^*$ that a specific BS can provide when the SINR is exactly equal to $\beta$. Note that the average data rate is here intended by referring to the geographical sub-areas, rather than the users located within any specific cell.

\vspace{-1mm} \subsection{Case studies}
This part of the paper discusses different realistic environments where WTBSs could be conveniently deployed, based on the simulation results we have obtained.
The first type of environment we propose is a rural area in a highly-developed country such as France, where both the telecom and the WT industries are very influential.
The next environment considered is a rural area in Ethiopia, where both the telecom and the WT industries are very limited but in rapid growth.
Finally we have an exurban environment in a windy area of Argentina, where only fewer WFs are available but there is a higher potential of power generation as well as a larger population needing connectivity.
For both cases, having considered the average wind speed density, we adopt useful open source data sets for population density \cite{fbDensity} and the locations of both CTs \cite{opencellid} and WTs \cite{data_Wpower,opsd}.
\par
All the simulations have been performed by considering the following system parameters:
$\alpha_L=2.2$ and $m_L=2$ for LoS transmission, $\alpha_N=3.2$ and $m_N=1$ for NLoS transmission, $\beta=-5$ dB, $\sigma_n^2=10^{-12}\,$W/Hz, $p_T=10\,$W and $h_T=30\,$m for CTs, and $p_W=11\,$W and $h_W=100\,$m for WTBSs.
We model the French and the Ethiopian case studies as rural environments and the Argentine one as a suburban environment according to \cite{al2014optimal}, that is:\\ 
(i) $\eta_3=-0.1\,$dB and $\eta_4=-21\,$dB respectively for 3G and 4G BSs, $\mathcal{S}_a=4.88$, and $\mathcal{S}_b=0.429$ for rural areas;\\
(ii) $\eta_3=-1\,$dB, $\eta_4=-20\,$dB, $\mathcal{S}_a=9.6117$, and $\mathcal{S}_b=0.1581$ when considering suburban environments.\\
Having assumed $R_3^*=2\,$Mbps and $R_4^*=17.5\,$Mbps respectively for 3G and 4G BSs, and biasing the association, all the simulation results have been averaged out over $10^4$ iterations, since the LoS or NLoS condition is a random variable.

\vspace{1.5mm}       \subsubsection{Western France}

\begin{figure*}
\centering
\includegraphics[width=0.9\columnwidth, trim={2.5cm 1.5cm 0cm 2cm},clip]{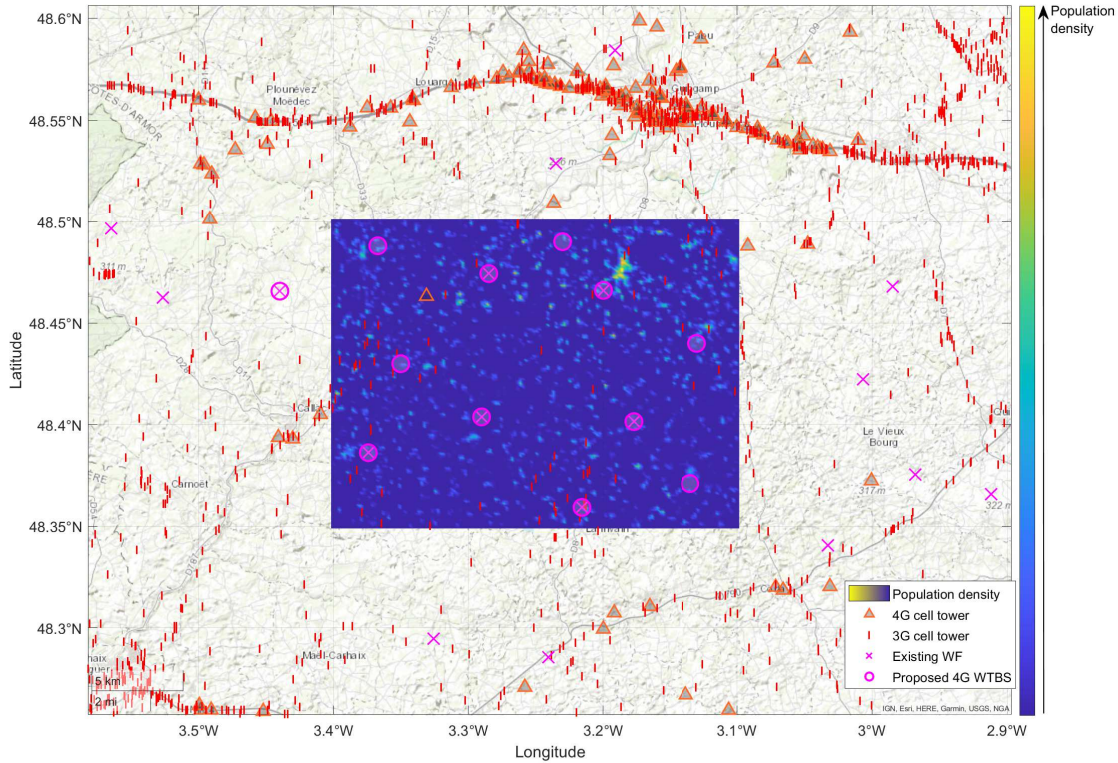}
\caption{System setup for the case study in western France.}
\label{fig:setup_wF}
\end{figure*}

\begin{figure*}
\centering
\subfloat[]{\includegraphics[width=0.5\columnwidth, trim={1.5cm 2cm 0cm 1.5cm},clip]{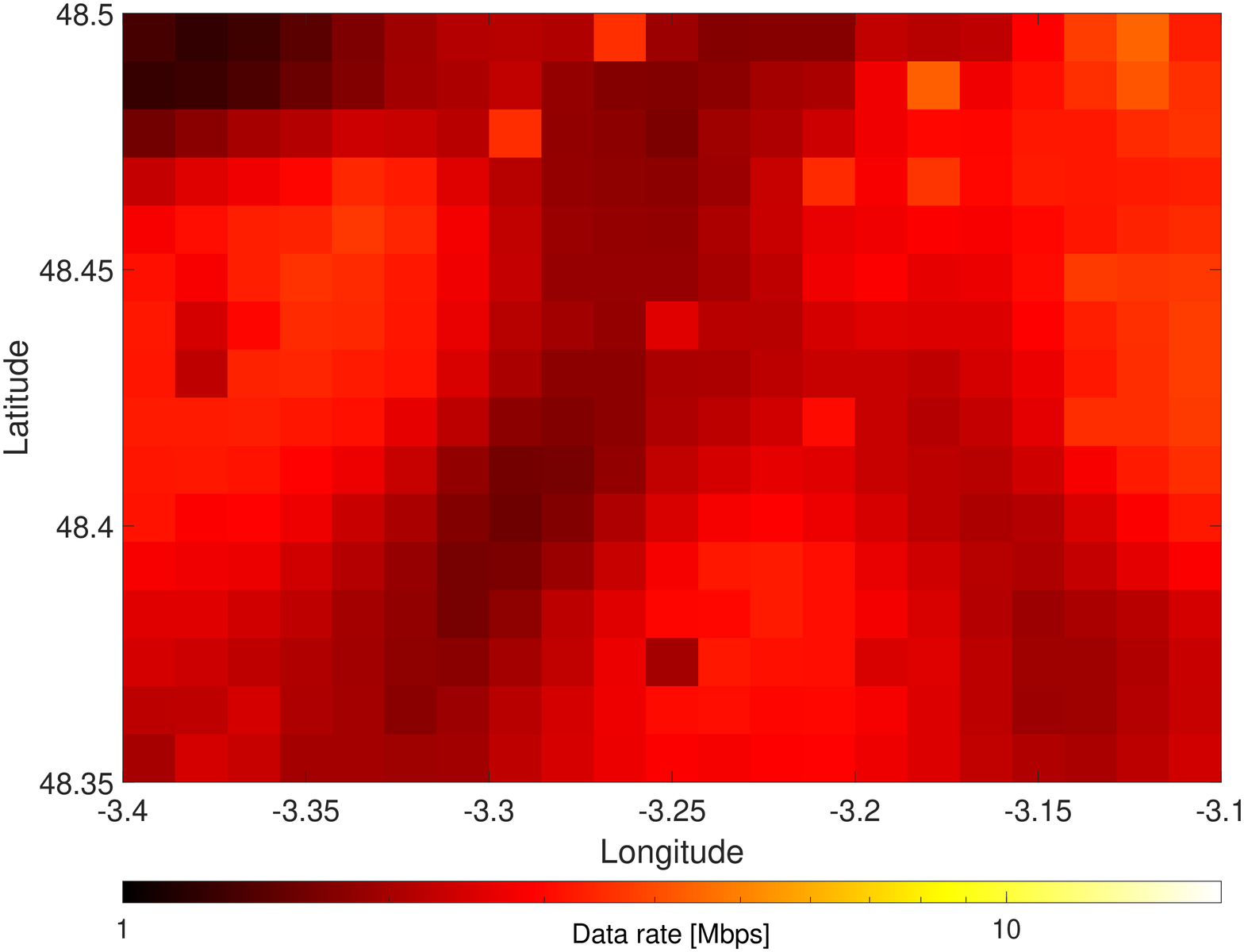} } 
\subfloat[]{\includegraphics[width=0.5\columnwidth, trim={1.5cm 2.05cm 0cm 1.5cm},clip]{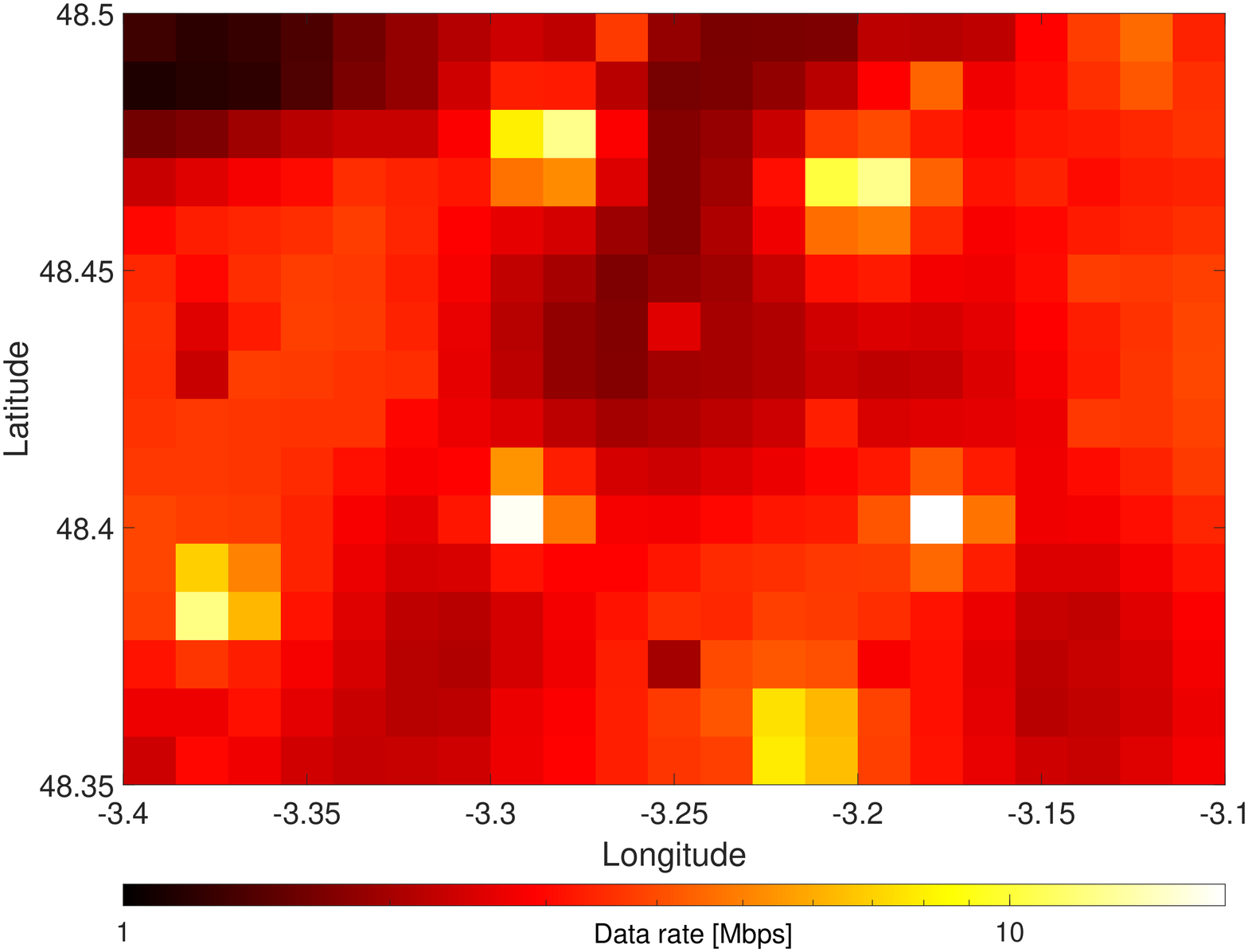} }\\
\subfloat[]{\includegraphics[width=0.54\columnwidth, trim={1.5cm 2cm 0cm 1.5cm},clip]{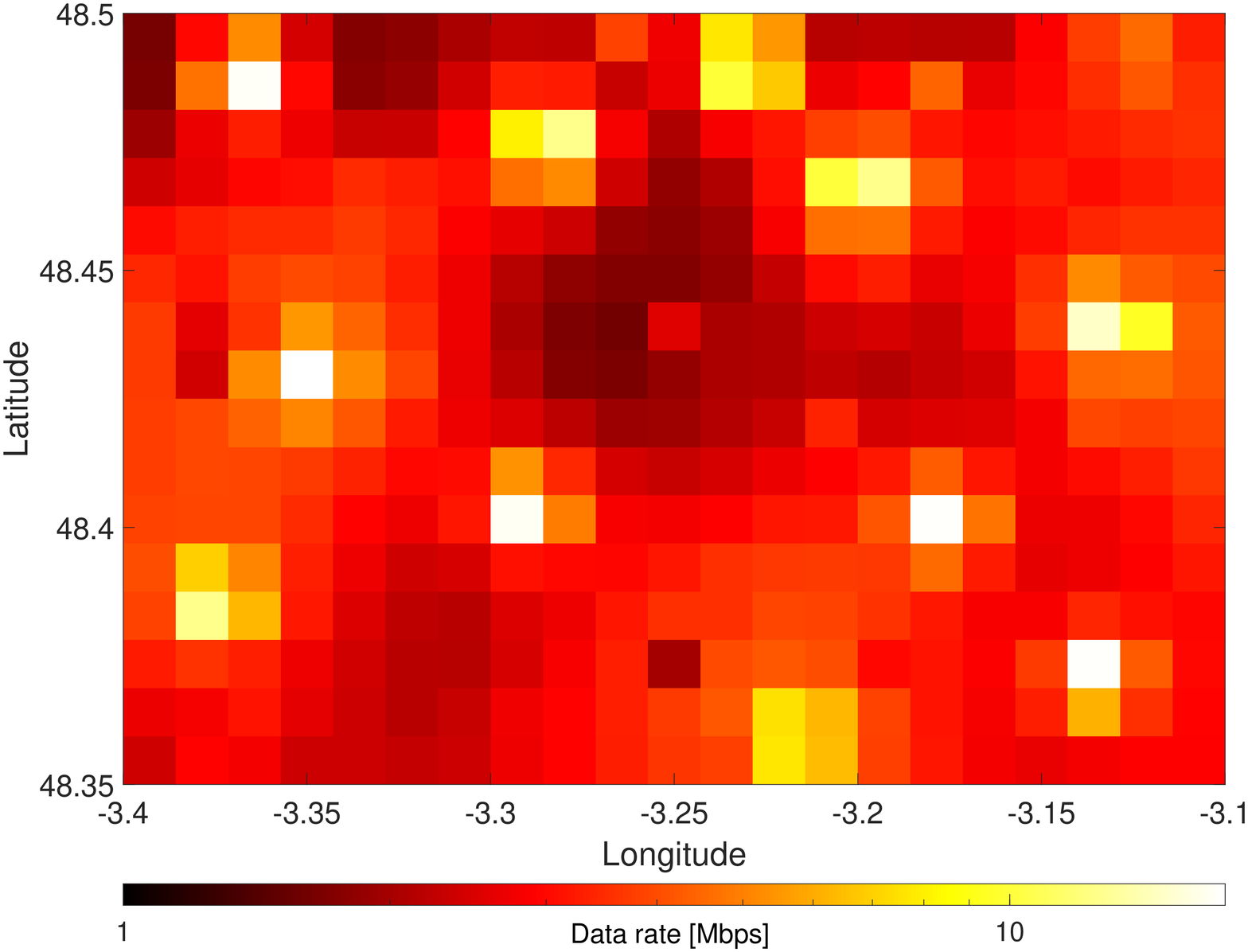} }
\caption {Data rate distributions in the western France case study when: (a) no WTBSs are deployed; (b) the proposed WTBSs are deployed; (c) both existing and proposed WTs are equipped.}
\label{fig:DR_wF}
\end{figure*}

The considered rural area is located between the towns Guingamp and Carhaix-Plouguer.
The area of interest occupies around $330\,$km$^2$, and its population density is depicted in Fig. \ref{fig:setup_wF}.
Here the yearly average wind speed almost equals $8\,$m/s \cite{GlobalWind} and large land surfaces would still be available for installing new WTs, although several ones are already present.
The existing TBSs are mainly providing 3G connectivity \cite{opencellid}, and the population is very sparse except for a small cluster on the North-East side that might be served by equipping an existing WT according to Fig. \ref{fig:setup_wF}. \par
In this case study, we evaluate the improvements that can be achieved by equipping all the existing WTs \cite{opsd} within and nearby the area of interest.
By optimizing the bias factor for associating to a 4G BS instead of a 3G one to a value of $29$, the current average data rate per user is $3.08\,$Mbps but could reach $4.12\,$Mbps by equipping the existing WTs.
However, Fig. \ref{fig:DR_wF} illustrates how some zones, such as the North-West side of the area, do not benefit from this solution due to their excessive distance from the additional BSs.
Actually, in this zone there are some areas in which the average data rate would be slightly affected due to the increase of the aggregate interference.
Thus, although the already promising results, it might be preferable to equip more of the existing WTs or even to build some new ones.
According to Fig. \ref{fig:DR_wF}c, indeed, the average data rate could considerably rise (up to $4.96\,$Mbps) by just adding five WTBSs.

\vspace{1.5mm}       \subsubsection{Central Ethiopia}

\begin{figure*}
\centering
\includegraphics[width=0.9\columnwidth, trim={1cm 0.7cm 0.6cm 1.8cm},clip]{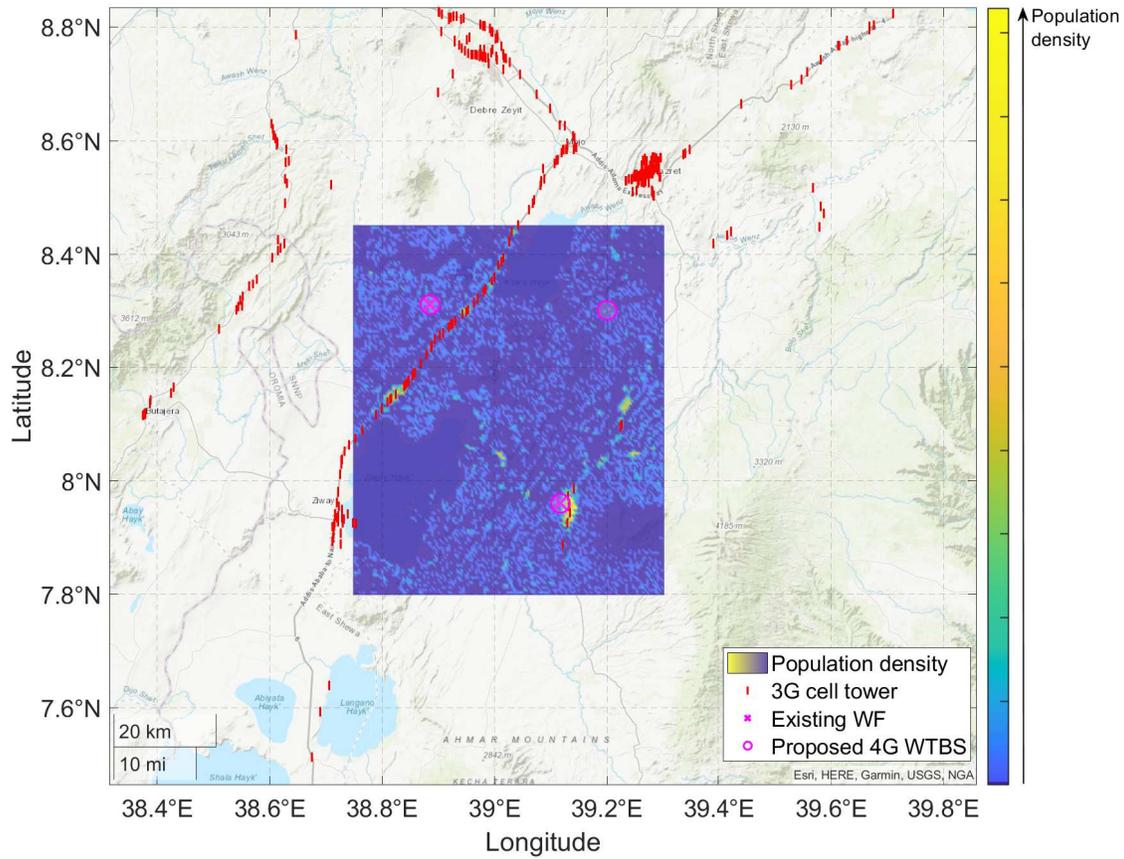}
\caption{System setup for the case study in central Ethiopia.}
\label{fig:setup_cE}
\end{figure*}

\begin{figure}
\centering
\subfloat[]{\includegraphics[width=0.46\columnwidth, trim={0.7cm 0.85cm 0.7cm 1.55cm},clip]{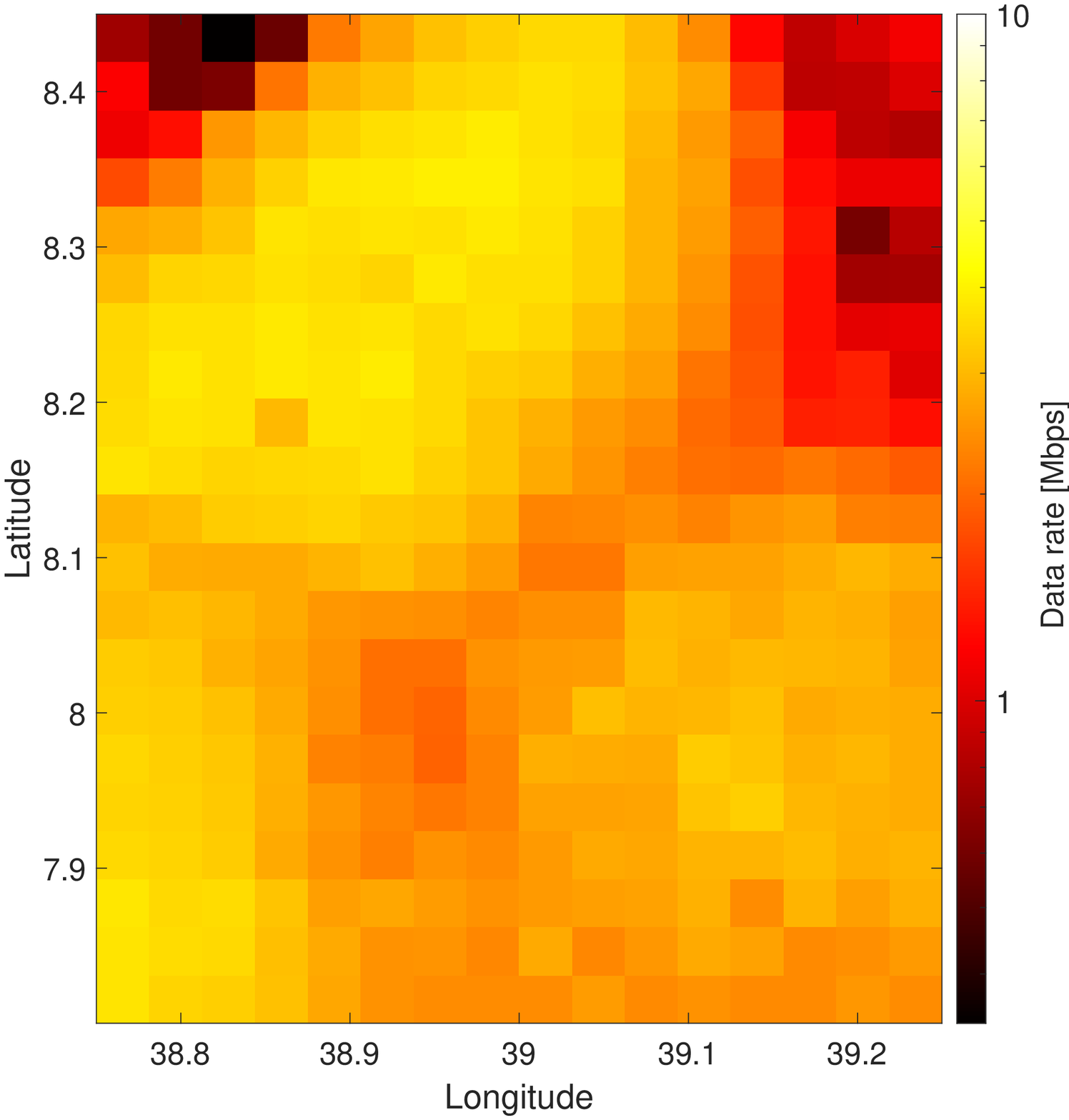} } 
\subfloat[]{\includegraphics[width=0.46\columnwidth, trim={0.7cm 0.75cm 0.7cm 1.55cm},clip]{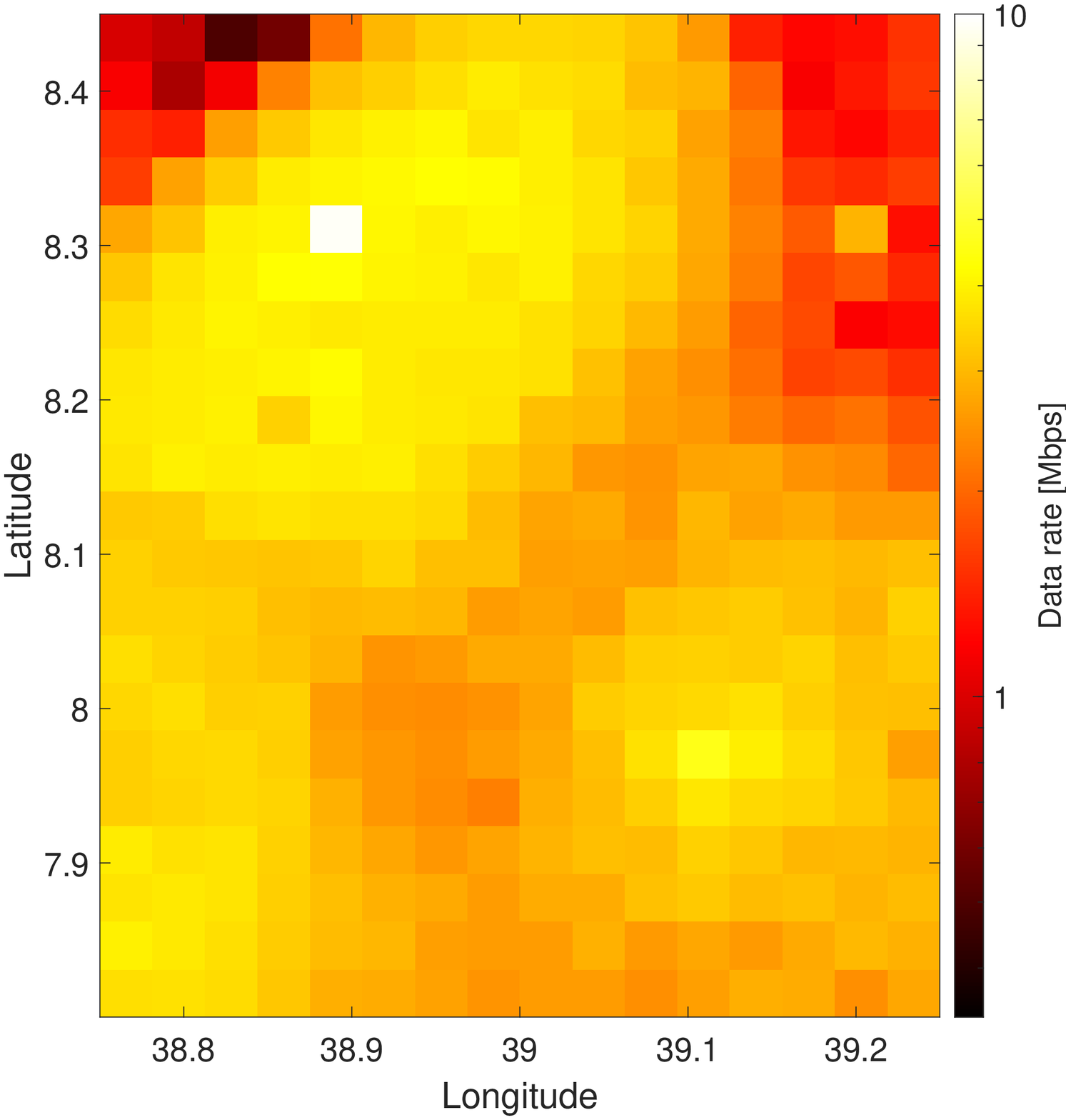} }
\caption {Data rate distributions in the central Ethiopia case study when: (a) no WTBSs are deployed; (b) the proposed WTBSs are deployed.}
\label{fig:DR_cE}
\end{figure}

Ethiopia is probably the most promising African countries for future installations of WFs. 
Moreover, this country needs to reduce drastically its imbalance index for telecom services \cite{zhang21telecommunication} in order to improve its economy, since almost its entire population cannot access 4G services. \par
The area of interest equals to roughly $4\,400\,$km$^2$ and hosts more than $91\,000$ inhabitants.
As displayed in Fig. \ref{fig:setup_cE}, this region is less than $100\,$km$^2$ far from the capital and it is still totally missing 4G CTs \cite{opencellid}. \par
Given the current presence of two WFs \cite{data_Wpower}, we propose to install an additional WTBS to cover the North-East corner of the area, which in general leads to a noticeable increase of $\bm\bar{R}$, as confirmed by Fig. \ref{fig:DR_cE}.
Starting from 2.88 Mbps, the average data rate per user has also been ameliorated by approximately $12\%$.

\vspace{1.5mm}       \subsubsection{Southern Argentina} 

\begin{figure}
\centering
\includegraphics[width=0.8\columnwidth, trim={0.3cm 2.3cm 0cm 1.5cm},clip]{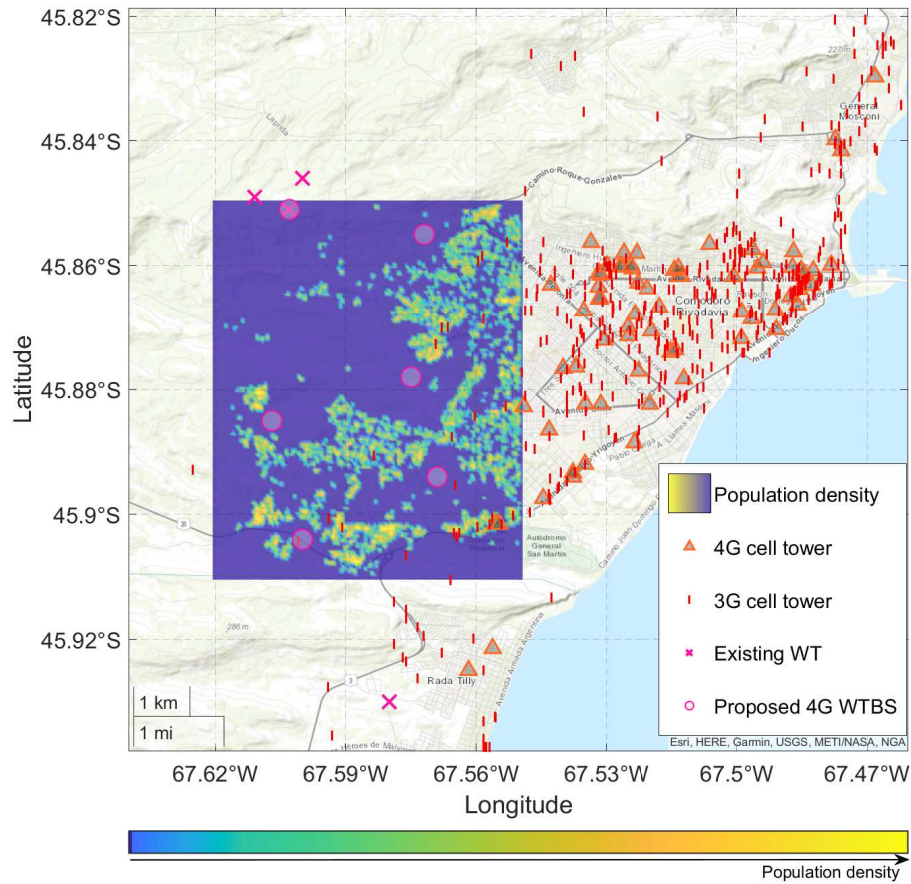}
\caption{System setup for the case study in southern Argentina.}
\label{fig:setup_sA}
\end{figure}

\begin{figure}
\centering
\subfloat[]{\includegraphics[width=0.46\columnwidth, trim={0cm 1cm 0cm 1.5cm},clip]{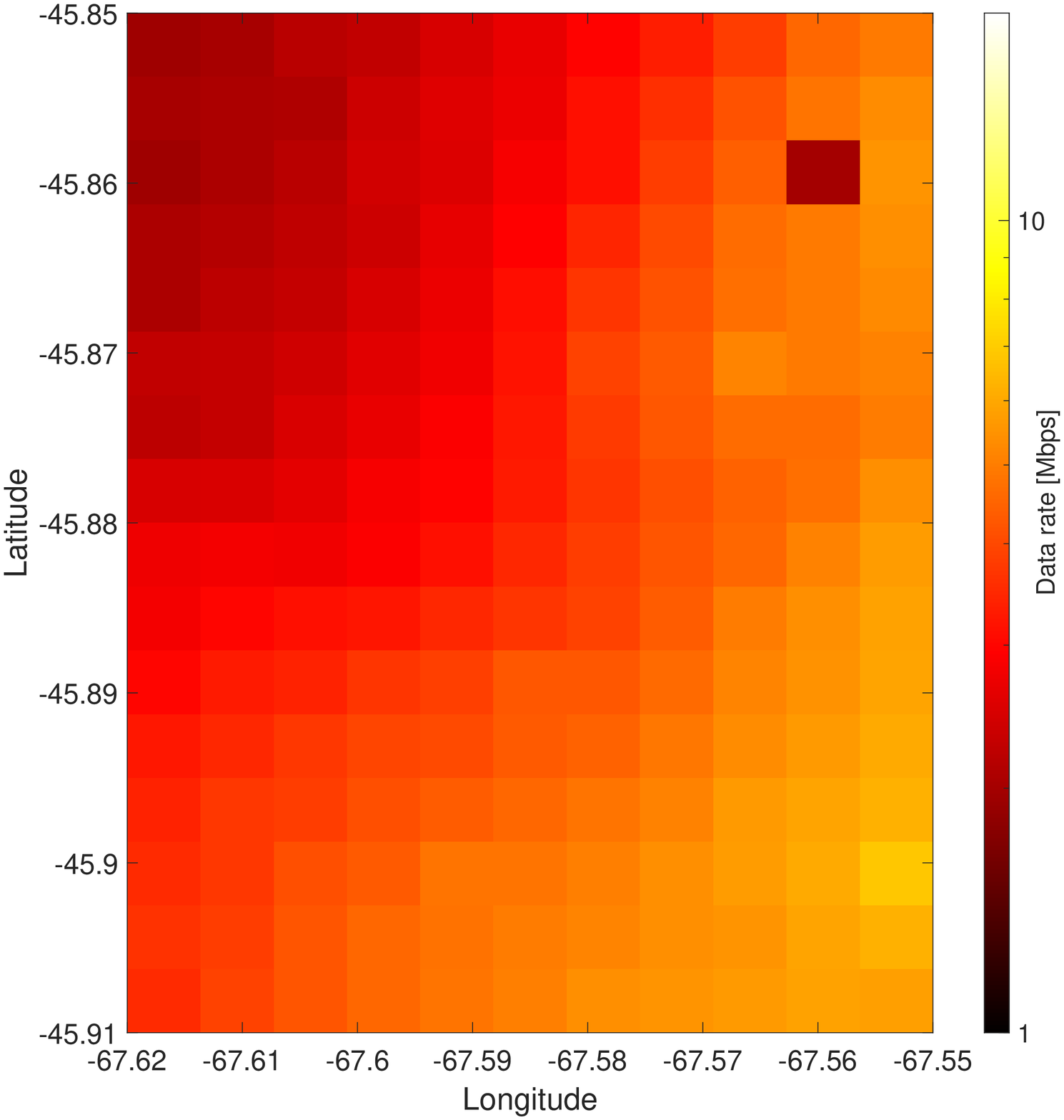} } 
\subfloat[]{\includegraphics[width=0.46\columnwidth, trim={0cm 1cm 0cm 1.5cm},clip]{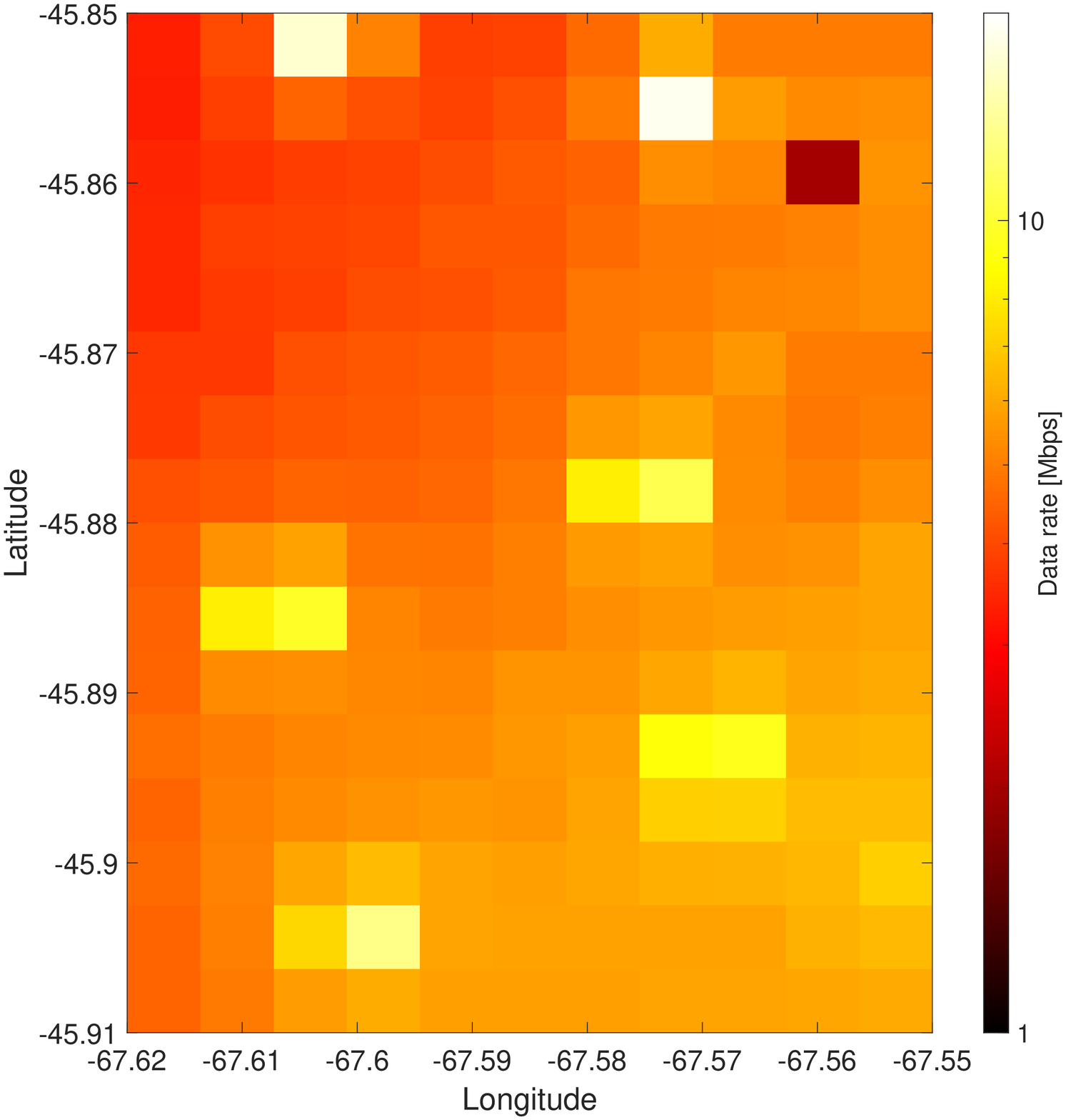} }
\caption {Data rate distributions in the southern Argentina case study when: (a) no WTBSs are deployed; (b) the proposed 4G WTBSs are deployed.}
\label{fig:DR_sA}
\end{figure}

The huge wind potential of this region has been partially exploited only few years ago. 
However, Argentina's installed capacity of WTs has almost tripled from 2018 to 2019 \cite{data_Wpower}, meaning that there is now a strong interest in this technology. \par
We have chosen a simulation area with approximately $27\,000$ inhabitants distributed over $20\,$km$^2$ in the western periphery of Comodoro Rivadavia.
This zone is characterized by a poor telecom infrastructure, according to \cite{opencellid}.
Since this town is gradually converting its power sources from oil to wind (the average speed is approximately $9.75\,$m/s \cite{GlobalWind}), we propose to deploy new WTBSs in some strategic locations, as illustrated in Fig. \ref{fig:setup_sA}. \par
Having optimized the bias factor to a value of $22$, the simulation results clearly show an overall improvement in terms of the average data rate available for the mobile users residing in this area.
Indeed, it can be inferred from the heat-maps in Fig. \ref{fig:DR_sA} that the number of proposed installations is sufficient to cover the entire area of interest.
Trivially, the zones with the highest average data rate are located in the proximity of the WTBSs, since they provide 4G connectivity and easily establish LoS links within distances of several hundred meters.
Note also that the areas far from any of these hybrid structures are not considerably affected by a significant increase in the interference.
Normalizing on the population density, the average data rate has been enhanced from $3.54\,$Mbps to $4.13\,$Mbps.

\vspace{1.5mm}       \subsubsection{Upcoming wind energy-based projects} 
Although the Kingdom of Saudi Arabia (KSA) is still almost solely relying on oil as an energy source, the city of NEOM is an ambitious project that expects to be fully powered by renewable energies.
According to the official website, wind and solar energies will also be used to produce green hydrogen, which can be better stored.
Several works have promoted the high wind potential of this region, which is characterized by high speed and slightly variable winds over the whole year.
A thorough case study has been conducted in \cite{WF_NEOM}, where a WF made of a hundred identical WTs (each one with rated power of $3.2\,$MW) has been proved to be profitable even without any government subsidy.
WTBSs would be thus an effective solution for limiting the visual impact on NEOM's landscape, especially given its goal of preserving $95\%$ of its natural resources (note also that off-shore WTs have a positive impact on sea life). 
\par
Another interesting project regards Denmark's largest construction ever, an energy island that is expected to produce up to $10\,$GW by mainly using off-shore WTs.
Evidently, the cost of the telecom infrastructure would be much higher in such a remote environment, apart from the fact that there is a very limited surface available.
Thus, off-shore WTs could be the perfect structures for hosting BS equipment, since they are bigger and stronger than their on-shore counterparts.  

\section{Conclusions and future work} \label{sec:conclusion}
By means of realistic case studies, we have showed that deploying WTBSs can be an effective solution for improving average data rate in various types of environments worldwide.
Therefore, we believe that WTs should be further incentivized in underdeveloped countries and rural areas in general.\par
An open problem might consist in optimizing the number and the locations of the WTBSs.
Future works should focus on the integration of this technology in a more complex network that includes also ABSs, LEO satellites, and aerial users, for instance.
Finally, the feasibility of mounting the BS equipment should be also evaluated for innovative and promising wind energy harvesting structures, such as bladeless wind oscillators, that aim to avoid problems of signal scattering or reflection due to the blades, acoustic pollution, and birds' death.

\bibliographystyle{IEEEtran}
\bibliography{wind}

\begin{thebibliography}{10}
\providecommand{\url}[1]{#1}
\csname url@samestyle\endcsname
\providecommand{\newblock}{\relax}
\providecommand{\bibinfo}[2]{#2}
\providecommand{\BIBentrySTDinterwordspacing}{\spaceskip=0pt\relax}
\providecommand{\BIBentryALTinterwordstretchfactor}{4}
\providecommand{\BIBentryALTinterwordspacing}{\spaceskip=\fontdimen2\font plus
\BIBentryALTinterwordstretchfactor\fontdimen3\font minus
  \fontdimen4\font\relax}
\providecommand{\BIBforeignlanguage}[2]{{%
\expandafter\ifx\csname l@#1\endcsname\relax
\typeout{** WARNING: IEEEtran.bst: No hyphenation pattern has been}%
\typeout{** loaded for the language `#1'. Using the pattern for}%
\typeout{** the default language instead.}%
\else
\language=\csname l@#1\endcsname
\fi
#2}}
\providecommand{\BIBdecl}{\relax}
\BIBdecl

\bibitem{Yaacoub19}
E.~{Yaacoub} and M.-S. {Alouini}, ``A key {6G} challenge and opportunity --
  {Connecting} the base of the pyramid: {A} survey on rural connectivity,''
  \emph{Proceedings of the IEEE}, vol. 108, no.~4, pp. 533--582, 2020.

\bibitem{zhang21telecommunication}
C.~Zhang, S.~Dang, B.~Shihada, and M.-S. Alouini, ``On telecommunication
  service imbalance and infrastructure resource deployment,'' 2021. Available
  online: https://arxiv.org/pdf/2104.03948.pdf.

\bibitem{Patent}
{Modification of wind turbines to contain communication signal functionality,
  by T. M. Sievert (Nov. 21, 2006). Patent US 7,138,961 B2 [Online]. Available:
  https://patents.google.com/patent/US20040232703A1/en}.

\bibitem{Matracia20rural}
M.~Matracia, M.~A. Kishk, and M.-S. Alouini, ``Coverage analysis for
  {UAV}-assisted cellular networks in rural areas,'' \emph{IEEE Open Journal of
  Vehicular Technology}, vol.~2, pp. 194--206, 2021.

\bibitem{Talgat2020}
A.~{Talgat}, M.~A. {Kishk}, and M.-S. {Alouini}, ``Stochastic geometry-based
  analysis of {LEO} satellite communication systems,'' \emph{IEEE
  Communications Letters}, 2020.

\bibitem{WindReview}
{Devashish}, A.~{Thakur}, S.~{Panigrahi}, and R.~R. {Behera}, ``A review on
  wind energy conversion system and enabling technology,'' in
  \emph{International Conference on Electrical Power and Energy Systems
  (ICEPES)}, 2016, pp. 527--532.

\bibitem{Ahmed14smartWF}
M.~A.~Ahmed and Y.-C. Kim, ``Communication network architectures for smart-wind
  power farms,'' \emph{Energies}, vol. vol. 7, pp. pp. 3900--3921, 06 2014.

\bibitem{FBsupercell}
A.~Tiwari, ``\em{SuperCell: Reaching new heights for wider connectivity}\rm,''
  2020. Facebook Engineering. \\ Available:
  https://engineering.fb.com/2020/12/03/connectivity/supercell-reaching-new-heights-for-wider-connectivity/.

\bibitem{WT_publicOpinion}
M.~Leiren, S.~Aakre, K.~Linnerud, T.~Julsrud, M.~Di~Nucci, and M.~Krug,
  ``Community acceptance of wind energy developments: Experience from wind
  energy scarce regions in {E}urope,'' \emph{Sustainability}, vol.~12, p. 1754,
  2020.

\bibitem{opencellid}
U.~labs, ``cell$\_$towers.csv.gz,'' Accessed: Dec. 8, 2020. [Online]. \\
  Available: https://www.opencellid.org/downloads.php.

\bibitem{data_Wpower}
{The Wind Power}, ``{Windfarms$\_$Argentina$\_$20210211.csv" and
  ``Windfarms$\_$Africa$\_$20210730.csv},'' Accessed: Aug. 12, 2021. [Online].
  Available:
  https://www.thewindpower.net/store$\_$windfarms$\_$view$\_$all$\_$en.php.

\bibitem{opsd}
{Open Power System Data (OPSD)}, ``renewable$\_$power$\_$plants$\_${FR}.csv,''
  Accessed: Nov. 25, 2020. [Online]. Available:\\
  https://data.open-power-system-data.org/renewable$\_$power$\_$plants/.

\bibitem{al2014optimal}
A.~Al-Hourani, S.~Kandeepan, and S.~Lardner, ``Optimal {LAP} altitude for
  maximum coverage,'' \emph{IEEE Wireless Communications Letters}, vol.~3,
  no.~6, pp. 569--572, 2014.

\bibitem{fbDensity}
{Humanitarian Data Exchange (HDX)}, ``population$\_$fra$\_$2019-07-01.csv.zip",
  "population$\_$eth$\_$2018-10-01.csv.zip" and
  ``population$\_$arg$\_$2018-10-01.csv.zip,'' Accessed: Nov. 4, 2020.
  [Online]. Available:
  https://data.humdata.org/search?groups=eth$\&$groups=arg$\&$q=population$\%$20\\density.

\bibitem{GlobalWind}
{Department of Wind Energy at the Technical University of Denmark, and World
  Bank Group. "Global Wind Atlas (GWA 3.0)". Accessed: Mar. 4, 2021. [Online].
  Available: https://globalwindatlas.info/}.

\bibitem{WF_NEOM}
F.~{Alfawzan}, J.~E. {Alleman}, and C.~R. {Rehmann}, ``{Wind energy assessment
  for NEOM city, Saudi Arabia},'' \emph{Energy Science $\&$ Engineering},
  vol.~8, pp. 755--767, 2020.

\end{thebibliography}

\vfill
\vspace{-50mm}

\begin{IEEEbiographynophoto}
{Maurilio Matracia}
[S'21] is a Ph.D. student at King Abdullah University of Science and Technology (KAUST). 
He received his B.Sc. and M.Sc. degrees in Energy and Electrical Engineering from the University of Palermo (UNIPA), Italy, in 2017 and 2019, respectively.
His main research interest is stochastic geometry, with special focus on rural and emergency communications.
\end{IEEEbiographynophoto}

\vspace{-10mm}

\begin{IEEEbiographynophoto}
{Mustafa A. Kishk}
[S'16, M'18] is a postdoctoral research fellow in the
communication theory lab at KAUST. 
He received his B.Sc. and M.Sc. degree from Cairo University in 2013 and 2015, respectively, and his Ph.D. degree from Virginia Tech in 2018.
His current research interests include stochastic geometry, energy harvesting wireless networks, UAV-enabled communication systems, and satellite communications.
\end{IEEEbiographynophoto}

\vspace{-10mm}

\begin{IEEEbiographynophoto}
{Mohamed-Slim Alouini} 
[S'94, M'98, SM'03, F'09] was born in Tunis, Tunisia.
He received the Ph.D. degree in Electrical Engineering from the California
Institute of Technology (Caltech), Pasadena, CA, USA, in 1998. 
He served as a faculty member in the University of Minnesota, Minneapolis, MN, USA, then in the Texas A$\&$M University at Qatar, Education City, Doha,
Qatar before joining KAUST, Thuwal, Makkah Province, Saudi Arabia as a Professor of Electrical Engineering in 2009. 
His current research interests include the modeling, design, and performance analysis of wireless communication systems.
\end{IEEEbiographynophoto}


\end{document}